\documentclass[conference,a4paper]{IEEEtran}
\usepackage{cite}
\ifCLASSINFOpdf
  \usepackage[pdftex]{graphicx}
   \graphicspath{{C:/Users/Administrator/Desktop/latex/eps}}
  \DeclareGraphicsExtensions{.pdf,.jpeg,.png,.eps}
\else
\fi
\usepackage{CJK}
\usepackage[numbers,sort&compress]{natbib}
\usepackage{cite}
\usepackage{amsmath,amssymb,enumerate}
\usepackage{cite,graphics,graphicx}
\usepackage{graphicx}
\usepackage{subfigure}
\usepackage{listings}
\usepackage{makeidx}
\usepackage{hyperref}
\usepackage{algorithm}
\usepackage{algorithmic}
\usepackage{float}
\usepackage{multirow}
\usepackage{amsmath}
\usepackage{amssymb}
\usepackage{xcolor}
\usepackage{epsfig}
\usepackage{graphics}
\usepackage{listings}
\usepackage{algorithm}
\usepackage{algorithmic}
\hypersetup{linkcolor=black, %
            citecolor=black, %
             pdftitle={Title}, %
            pdfauthor={Name}, %
           pdfsubject={Subject}, %
          pdfkeywords={Key words}}
\begin{document}

\title { Promote the Industry Standard of Smart Home in China by Intelligent Router Technology }

\author{\IEEEauthorblockN{Hui Lin\IEEEauthorrefmark{1},
Jianbiao Lin\IEEEauthorrefmark{1},
Ke Ji\IEEEauthorrefmark{2},
Jingjie Wang\IEEEauthorrefmark{3},
Feng Lin\IEEEauthorrefmark{4}}
\IEEEauthorblockA{
Department of Computer Science and Engineering\\
Sichuan Unversity, Jinjiang College, Pengshan, Sichuan, 620860, China\\
Colege of Civil Engineering\\
Sichuan Unversity, Jinjiang College, Pengshan, Sichuan, 620860, China\\
zeroyuebai@hotmail.com}
}


%Telephone: (800) 555--1212, Fax: (888) 555--1212}


\maketitle

\begin{abstract}
%\boldmath
 The reason why smart home remains not popularized lies in bad product user experience, purchasing cost, and compatibility, and a lack of industry standard\cite{avgerinakis2013recognition}. Echoing problems above, and having relentless devoted to software and hardware innovation and practice, we have independently developed a set of solution which is based on innovation and integration of router technology, mobile Internet technology, Internet of things technology, communication technology, digital-to-analog conversion and codec technology, and P2P technology among others. We have also established relevant protocols (without the application of protocols abroad). By doing this, we managed to establish a system with low and moderate price, superior performance, all-inclusive functions, easy installation, convenient portability, real-time reliability, security encryption, and the capability to manage home furnitures in an intelligent way. Only a new smart home system like this can inject new idea and energy into smart home industry and thus vigorously promote the establishment of smart home industry standard.   \\

%\bibliographystyle{IEEEtran}
%\bibliography{IEEEabrv,bare_conf}
\end{abstract}


\section{Introduction}

Since this year, the waves of smart home are on its rise. This October, a company that sells intelligence temperature controllers NEST of Google acquired Revolv, a smart home central control equipment start-up. Xiaomi released its smart plug, smart camera, among four smart end new products. Enterprises at home and abroad one after another plunge into the big cake of smart home. Smart home is opening new vista and space for Internet and household appliance industry. All View Consulting forecasts that by 2020, the ecological product of domestic smart home appliance in China will reach one trillion yuan. SAIF Partners predict that the scale of smart home industry by traditional definition in China will reach 5.5 billion yuan in 2014, and that number will soar to 7.5 billion yuan in 2015. However, three stumbling blocks are in the advancing way of smart home industry: user experience, purchasing cost, and lousy compatibility\cite{albuquerque2014solution}. In response, we have independently developed a set of solution based on smart router technology. There are many problems existing in routers in the market. Firstly, wireless control based on 315M, 433M and other frequency ranges has no network protocol and can only send simple control command. Collision occurs when there are over three connected devices, which renders the process more difficult to succeed. Secondly, control network based on ZigBee registers small range, poor through-the-wall performance, complex protocol, inordinate price, and at the same time is exclusive and incompatible to devices existing in the market. The third one is control network based on WI-FI. WI-FI boasts a small control range and thus is limited to only a few connected devices. Normally when household router is connected to over ten devices the network would drop or other instabilities would happen. Having taken characteristics above into consideration, echoing the needs for transmission distance, stability, and controlled quantity, our system has utilized control network based on 433M frequency independently developed a self-organized protocol based on the control network which could bear dynamic networking functions similar to that of Zigbee and boast a connected devices number of over 100. As a result, our system has managed the networking capabilities of Zigbee with high connected devices number and long distance transmission. Also the protocol of 433M frequency is an open one and can be compatible with smart devices currently in the market. Therefore, smart router will change the landscape of smart home market and high-end router market and lays a solid foundation for establishment of industry standard for China smart home industry.

\section{A set of solution aiming at promoting establishment of industry standard for China smart home market}\label{SEC: A set of solution aiming at promoting establishment of industry standard for China smart home market}

\subsection{The set of solution includes}\label{SSEC: The set of solution includes}

Smart router,cloud server, mobile terminal, and intelligent terminal. Meanwhile, the system is a intelligence development platform which enables programmers worldwide to carry out secondary development on this platform and thus an ecological chain with sound circulation takes shape (similar to AppStore of Apple).

\subsection{Feature of the whole set of solution}\label{SSEC: Feature of the whole set of solution}

As a household smart center, the smart router is able to administer all of the connected devices, control smart devices in houses, keep houses in security under surveillance, monitor household environment in terms of temperature, humidity, PM 2.5 etc., alarm the police when household accidents happen such as smog or gas, enable users to be remotely connected to the smart center through devices such as cell phone and PAD at any time anywhere to observe and supervise household appliances, and at the same time promptly watch surveillance video in the house. To speak of, the cost of this device is only a small percentage of a traditional smart home product.

Users can know at first hand the environment of household through cell phones in many ways such as gas leakage, burglar break-in, illegal opening of doors or windows, abnormal temperature and humidity, smoke alarm, touching of valuable things, real-time temperature and humidity detection, PM2.5 detection. It can be seen as a safety housekeeper. At the same time, users can be promptly aware of the surveillance video image in the house through cell phones, which could provide a more perceptual and intuitive supervision to surrounding environment. Moreover, the router can control household appliances such as television, air conditioner as well as all the other household appliances with remote control. Last but not least, the router is also a cloud service center where users can put personal data in family cloud.

\subsection{Introduction to specific functions}\label{SSEC: Introduction to specific functions}

\subsubsection{Functions of router}\label{SSSEC: Functions of router}

Just like other ordinary routers, the router can visit the Internet and distribute WI-FI data\cite{zualkernan2009infopods}.

\subsubsection{Intellisense}\label{SSSEC: Intellisense}

Users can receive alarm in houses through cell phones. Smart router can be adaptive to every alarm apparatus. When there is an alarm from apparatuses, smart router will promptly send alarm information to cell phones of users and provide security for users. Alarm apparatus spans gas leakage, burglar break-in, illegal opening of doors or windows, abnormal temperature and humidity, smoke alarm, touching of valuable things. Meanwhile, users can promptly get to know the temperature, humidity, and PM2.5 of houses.

\subsubsection{Intelligent surveillance}\label{SSSEC: Intelligent surveillance}

Smart router can be installed with USB camera of low price as well as wireless camera, which is convenient for users to obtain real-time image through cell phones. Wireless camera can be based on codec of H264, which makes image clear and smooth.

\subsubsection{Intelligent plug}\label{SSSEC: Intelligent plug}

Cell phones can remotely control the on and off of plug, which means controlling the appliances connected to the plug.
\subsubsection{Intelligent cloud service}\label{SSSEC: Intelligent cloud service}

Smart router can serve as a cloud service center that provides personal data management for family members and boasts a good level of privacy.

\subsubsection{Intelligence remote control}\label{SSSEC: Intelligence remote control}

X-Router enables users to put aside all remote controls at home and control all the household appliances through X-Router. For example, users can control lights, curtains, plugs, television, air conditioners, DVD, and STB through cell phones.

\subsection{Main technologies}\label{SSEC: Main technologies}

1) Communication protocol to control establishment and implementation of protocol.

2) Transfer function of communication protocol to achieve intelligence transfer of transmission and P2P.

3) P2P technological research to achieve P2P of low flow.

4) Establishment and implementation of camera protocol with original server protocol integrated.

5) development of cell phone software and server side software. Establishment and implementation of chat protocol between software.

6) Device terminal networking.

7) establishment and implementation of connection control protocol of device terminal and router.

8) Establishment and implementation of control protocol of device terminal of each types. Protocols for different types of devices are different and are individually carried out.

9) Development and production of communication printed circuit board of communication device terminal.

10)  Build software and hardware platform for routers.

11) Printed circuit board hardware circuit diagram design with low consumption, high simultaneous access, and long duration.

\subsection{Technical index}\label{SSEC: Technical index}

\subsubsection{Low consumption}\label{SSSEC: Low consumption}

the transmitted power is only around 1mW. It also uses sleep mode with low power dissipation, making the device use much less electricity. According to estimates, the device can endure a continuous, active period of six months to two years just by two AA batteries, which other wireless devices can hardly match.

\subsubsection{Low cost}\label{SSSEC: Low cost}

the cost for the whole set of solution is around 100 yuan.

\subsubsection{Short time delay}\label{SSSEC: Short time delay}

communication delay and delay period for activation from sleep mode are both very short. Delay for a typical search equipment is 30ms, 15ms for activation from sleep mode, 15ms for active devices to join via channels. Thus the technology is best suited for application in wireless control that is highly commanding in time delay (such as industry control).

\subsubsection{Self-organized network technique}\label{SSSEC:Self-organized network technique}

a star schema has the maximum capacity for 254 slave units and one primary device. And the network is flexible.

\subsubsection{Reliability}\label{SSSEC: Reliability}

 Strategy to avoid collision is employed. Specialized slot time is conserved for communication business in need of stable bandwidth, so that competition and conflict for sending data are avoided. The MAC layer employs a completely definite data transmission mode where each sent data package must wait for the confirmation from the recipient. If any problem occurs in the transmission process, the data can be sent again.

\subsubsection{Security}\label{SSSEC: Security}

HTTPS encryption that supports authentication and certification and uses encrypted algorithm.

\subsubsection{P2P technology}\label{SSSEC: P2P technology}

a technology that deals with NAT gateway or firewall penetration.

\subsubsection{Low power dissipation technology of cell phone clients}\label{SSSEC: Low power dissipation technology of cell phone clients}

 a real-time technology that researches low power dissipation.

\subsection{Explanation of key technologies of each terminal}\label{SSEC: Explanation of key technologies of each terminal}

\subsubsection{Server terminal}\label{SSSEC: Server terminal}

Server terminal is the bridge for the whole system to connect where cell phones and routers build data and exchange data, and carry out P2P (peer-to-peer) communication. The burden of servers is reduced. Meanwhile servers provide login and registration as well as user information management for users. Servers can expand and form a cluster according to workload dynamic, thus increasing the processing capacity for cell phones and routers\cite{min2013design}.

We have pulled up a protocol of our own, which could identify which type of connection employed judging the type of request (Normally small data is transmitted through server UDP, and large data uses P2P connection). It connects the data path from cell phones to routers and make it possible to transparently transmit data from cell phones to router terminals.

\subsubsection{Smart router terminal}\label{SSSEC: Smart router terminal}

Smart router terminal is the core part of our project, on which we build software and hard ware platform of routers. Firstly, it is a router with a high level of performance by which common PC and cell phones can visit the Internet. Secondly, it is our data processing center that achieves our core interaction protocol, and code and decode all the control alarm data and then process them. Thirdly, its radio frequency identification function enables it to send and receive radio frequency,and use the protocol we develop to code and decode to process digital signal and analog signal.

By virtue of the protocol, it connects the data path from routers to device terminals. With the aid of servers, it makes possible transparent transmission of data from cell phones to device terminals.

\subsubsection{Cell phone}\label{SSSEC: Cell phone}

As long as cell phones of users can be connected to 2G/3G/4G/wifi Internet, users can be conveniently connected to servers via software of cell phones, transparently communicate with smart routers and device terminals through the protocol.

Users can register by cell phone software and log on the server, modify their personal information by HTTP communication protocol, add friends by chat protocol, and chat with friends.
Cell phones can send router administer command through (the path between cell phones and routers) introduced above and thus set up for switches of routers, wireless switches, and PPPOE.

When cell phones terminal want to control device terminal, a control protocol will be generated and be sent to the device terminal through the (path from cell phone to device terminal) introduced above. The device terminal receives the decoded information of the control protocol and obtains the specific command and implement it. Device terminal needs to report its state or generate the corresponding reporting command and send it to cell phones terminal through the path introduced above when it receives the alarm information which needs to be reported to users. The cell phones terminal receives the reported decoded protocol, obtains the specific command, and then registers its state on the UI of the application software or alarms the police.

\subsubsection{Device terminal}\label{SSSEC: Device terminal}

Device terminal in effect realizes the communication protocol of the solution and can be seamlessly and smoothly extended. Any protocol that realizes the solution can join the router and be part of the networking to control the router.

Device terminal spans all the alarm apparatuses, plug, bulb, power supply, sensor, and household appliance.

\subsubsection{Termianl SDK}\label{SSSEC:Termianl SDK}

If we can connect data channels between each device, we can do a lot of things on it. Our solution is open to SDK of device communication. By virtue of SDK, the third party can control devices that it develops through its application and achieve platformization purpose.

\subsubsection{6)Camera cloud serive cluster and camera}\label{SSSEC: 6)Camera cloud serive cluster and camera}

Considering features of camera such as strong performance of processor, large data throughput, and large amount of data of images, as we need to ensure that camera does not affect the stability of other devices\cite{yuan2014accountability}, we make our camera separate image data and control data. Transmission and storage of image data are solely handled by server clusters.

Communication of camera terminal and cell phones can be realized by server transfer, or by P2P connection through hole punching technique of servers and cell phones.

When cell phone terminals need to be connected to cameras, they will also be connected to camera cloud service cluster. They firstly make a request for P2P, if the path does not support it, it will commence server transfer. Control operation of cell phones such as rotating the camera, shifting up, down, to left, or right, is carried out through the path to the main servers. Cell phone terminals can local-save image date of cameras, or directly save the data on remote Yunfile such as Yun Baidu and Kuaipan via users' Yunfile accounts.

\subsection{Content of innovation technology}\label{SSEC: Content of innovation technology}

\subsubsection{Combination of radio frequency technology and router}\label{SSSEC: Combination of radio frequency technology and router}

Tradition home gateway master control systems all employ specialized processors and communication protocol abroad with only single function and inordinate price\cite{nejad2014operation}. With the development of chip technology, the function of chips of household routers are powerful enough to carry out control over smart home appliances\cite{razminia2011chaotic}. Therefore, we design this smart router, add radio frequency module to routers, write programmes for control and communication protocol, thus achieve a master control system with multi-functions, low cost, and high stability, which can also be used as general routers.

\subsubsection{Home gateway communication system}\label{SSSEC:Home gateway communication system}

Home gateway communication system utilizes exclusive custom protocol which can not be compatible with third party system and makes it difficult to upgrade\cite{garcia2012smart}. The system hopes to maintain open and universal in its application and integrates elements of Jingle protocol used in social network. The system also adds certification, control, and device discovery functions to the protocol, thus the system is robust in its expansiveness, can interact with any server that supports Jingle protocol, and communicate with any client that supports Jingle protocol. Meanwhile, the system supports device control and social interaction. As a result, our system can not only control smart devices, but also enable family members to communicate via social network.

\subsubsection{3£©Control network protocol}\label{SSSEC:3£©Control network protocol}

There are three types of control network by traditional definition:
The first one is wireless control based on 315M, 433M and other frequency ranges. The benefit of such control network is a high level of accuracy and a far-reaching control range as much as 1000 meters if power reaches its maximum. On the flip side, there is no network protocol and it can only send simple control command. Collision occurs when there are over three connected devices, which renders the process more difficult to succeed\cite{han2014generating}.

The second one is control network based on ZigBee. Its shortcomings lie in small range,poor through-the-wall performance, complex protocol,inordinate price,and at the same time it is exclusive and incompatible to devices existing in the market\cite{garcia2013multi}.

The third one is control network based on WI-FI. WI-FI boasts a small control range and thus is limited to only a few connected devices. Normally when household router is connected to over ten devices the network would drop or other instabilities would happen\cite{kammerer2012router}.
Having taken characteristics above into consideration, echoing the needs for transmission distance, stability, and controlled quantity, our system has utilized control network based on 433M frequency independently developed a self-organized protocol based on the control network which could bear dynamic networking functions similar to that of Zigbee and boast a connected devices number of over 100.  As a result, our system has managed the networking capabilities of Zigbee with high connected devices number and long distance transmission. Also the protocol of 433M frequency is an open one and can be compatible with smart devices currently in the market, such as 433M infrared body detecting alarm.

\subsubsection{High-speed router}\label{SSSEC:High-speed router}

The transmission rate of traditional router is only 150m/300M. Our router employs AC technology and thus the transmission rate can reach as much as 900M.

\begin{figure}[!htb]
    \centering
    \includegraphics[height=0.29\textwidth,width=0.50\textwidth]{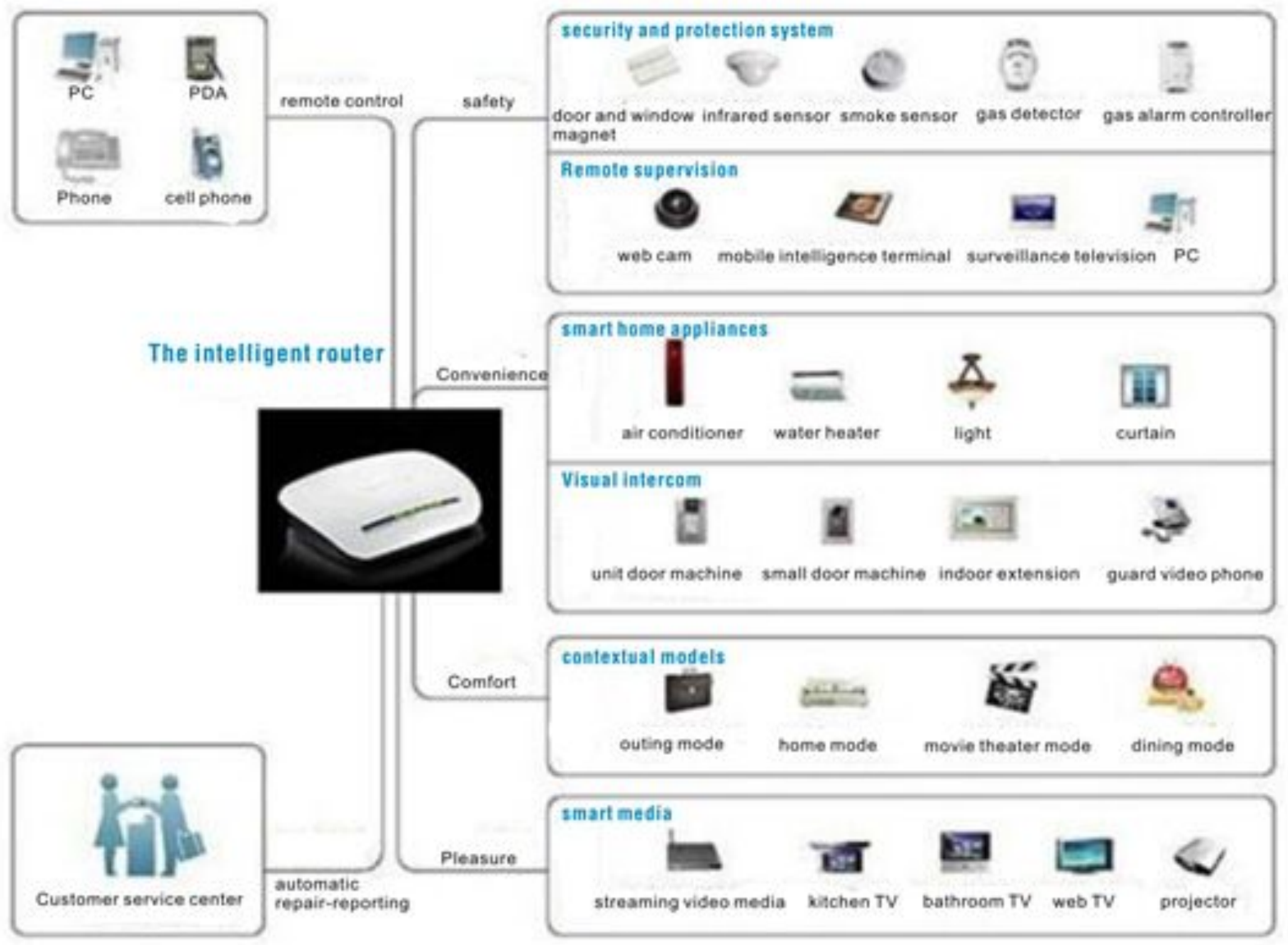}\\
    \caption{\label{fig:2-1}Product concept structure graph.}
\end{figure}

\subsection{Theoretical basis (including data transmission)}\label{SSEC:Theoretical basis (including data transmission)}

\subsubsection{TCP/IP protocol}\label{SSSEC:TCP/IP protocol}

TCP/IP protocol is the abbreviation for Transmission Control Protocol/Internet Protocol. It is also known as network communication protocol. It is the fundamental protocol of Internet and the foundation of international Internet network, comprised of IP protocol at network layer and TCP protocol at transport layer. TCP/IP defines the standard for how electrical devices are connected to Internet and how data transmits between them. The protocol employs a hierarchical structure of four layers and each layer calls the protocol provided by its following layer to fulfil its need. To put it in blank words, TCP is in charge of spotting problems occurring in transmission and once problem occurs it sends our signals to command a new transmission until all the data is safely and correctly sent to the destination. While IP set an address for each connected device of Internet.

The communication of all the terminals and devices in this project is based on TCP/IP protocol.

\subsubsection{Jingle/XMPP protocol}\label{SSSEC:Jingle/XMPP protocol}

XMPP(Extensible Messaging Presence Protocol) is a protocol based on extensible markup language (XML) applied in instant messages (IM) and online presence detection.It promotes the on-time and instant operation between servers. The protocol is likely to ultimately allow Internet users to send instant messages to others on the Internet even if the operating systems and browsers are different.

The P2P connection and transmission in this solution are both achieved based on revised Jingle XMPP protocol.

\subsection{Experimental basis}\label{SSEC:Experimental basis}

The router structure as follows is the ultimate version utilized by this project after multiple practices.

\begin{figure}[!htb]
    \centering
    \includegraphics[height=0.29\textwidth,width=0.50\textwidth]{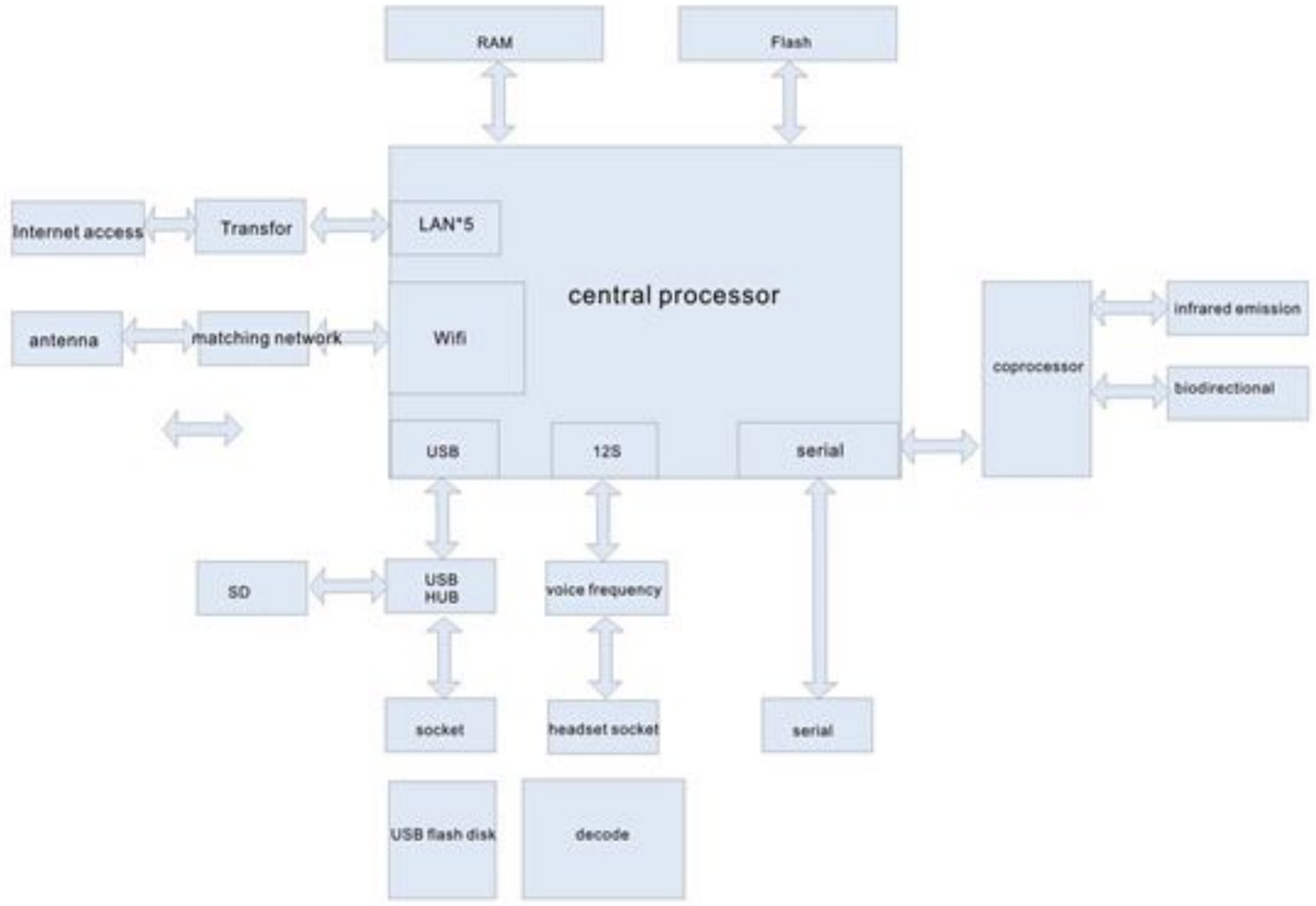}\\
    \caption{\label{fig:3-1}The Router Frame.}
\end{figure}

\subsection{Product index}\label{SSEC:Product index}
\subsubsection{Smart route}\label{SSSEC:Smart route}  \textrm{}

1)size: 23£ª23£ª5

2)transmission standard: IEEE 802.11ac/b/g/n

3)wireless transmission rate: 300M+700M

4)antenna gain: 5dbi

5)wired Internet access: one WAN, four LAN, 1000/100/10subject to adjustment and configuration

6)safety standard: 64/128 WEP encryption technology, WPA/WPA2 encryption, supports WPS WI-FI Protected Setup function

7)transmission band: 2.4GHz and 5.0GHz

8)reset key: 1

9)WPS key: 1

10)supported connected device terminals: 255

11)capable of responding to command of cell phone clients and each device terminals in three seconds.

\subsubsection{Server}\label{SSSEC:Server}  \textrm{}

1)One server can simultaneously support 6000 users.

2)7*24h smooth operation

\subsubsection{Wireless device}\label{SSSEC:Wireless device}  \textrm{}

1)effective range: outdoors 1500M£¬indoors 300M

2)frequency range: 240-930MHz

3)FSK, GFSK and OOK modulation mode

4)maximum output power: 20dBm

5)sensitivity: -121dBm

6)low power dissipation: 18.5mA (reception); 85mA@+20dBm (transmission)

7)data transmission rate: 0.123-256kbps

\section{Conclusion}\label{SEC: Conclusion}

After relentless research and practice, our final smart router has greatly overcame the three big shortcomings of smart home, user experience, purchasing cost, and bad compatibility. We bring a revolution to the smart home market and make it accessible to common people. We enable common people to experience a life of intelligence and high quality with a moderate price, thus making smart home more popularized in China market.

The appearance of smart routers also brings new vista to the smart home market and high-end router market, bring fresh vibrancy and business opportunities to China market. Also industry standard of China smart home will achieve breakthroughs with further research and development of smart router technology.

In researching and developing the technology of smart router, we have come across some inevitable problems. For example, during research and development, although we harbour a strong awareness of protection to the environment and carried out protection measures, the external environment is still affected. But we believe that when products are mature enough to be massively produced, relevant technologies will surely be applied in avoiding environment pollution risk from the source or reducing it.

\section{ Acknowledgement}\label{SEC:  Acknowledgement}

The research subject was supported by the department of Civil Engineering and the department of Computer Science$\&$Engineering Jinjiang College, Sichuan University.  I would like to express my thanks to Prof. Bingfa Lee¡¯s suggestions and guidance, as well as Guanguuan Yang$\&$Zhuo Li and Hui Zhang whose books give me a lot of inspiration.





%\bibliographystyle{IEEEtran}
\fontsize{8pt}{\baselineskip}\selectfont
\bibliographystyle{IEEEtran}

\bibliography{GG}


% \fi

\end{document}